\documentstyle[aps,prl,epsfig,preprint,floats]{revtex}
\begin{document}
\epsfysize3cm
\epsfbox{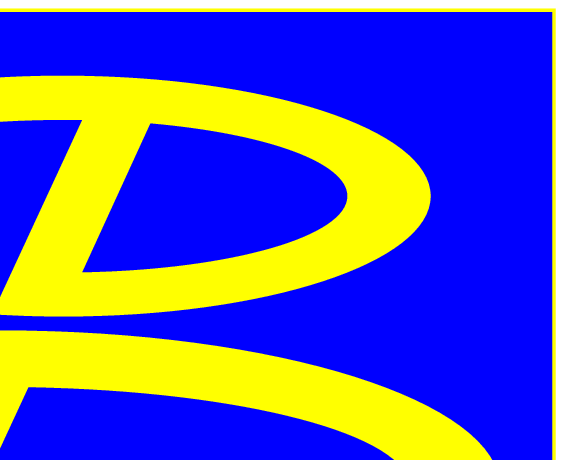}

\vskip -3cm
\noindent
\hspace*{12cm}KEK Preprint 2001-34 \\
\hspace*{12cm}Belle Preprint 2001-9 \\

\vskip 1cm

\renewcommand{\baselinestretch}{1.2}

%

\begin{center}
{\Large\bf
 Search for Direct CP Violation in $B \rightarrow K\pi$
Decays
\footnote{Submitted to PRD Rapid Communications.}}
\vskip 0.5cm
(The Belle Collaboration)
\vskip 0.2cm
{\normalsize
  K.~Abe$^{9}$,               
  K.~Abe$^{37}$,              
  R.~Abe$^{27}$,              
  I.~Adachi$^{9}$,            
  Byoung~Sup~Ahn$^{15}$,      
  H.~Aihara$^{39}$,           
  M.~Akatsu$^{20}$,           
  Y.~Asano$^{44}$,            
  T.~Aso$^{43}$,              
  V.~Aulchenko$^{2}$
% ...
}         
\end{center}
\small
\begin{center}
% ...
$^{2}${Budker Institute of Nuclear Physics, Novosibirsk}\\
% ...
$^{9}${High Energy Accelerator Research Organization (KEK), Tsukuba}\\
% ...
$^{15}${Korea University, Seoul}\\
% ...
$^{20}${Nagoya University, Nagoya}\\
% ...
$^{27}${Niigata University, Niigata}\\
% ...
$^{37}${Tohoku Gakuin University, Tagajo}\\
% ...
$^{39}${University of Tokyo, Tokyo}\\
% ...
$^{43}${Toyama National College of Maritime Technology, Toyama}\\
$^{44}${University of Tsukuba, Tsukuba}\\
% ...
\end{center}
\normalsize

\maketitle
%
%
\begin{abstract}
We search for direct CP violation in flavor specific $B\rightarrow K\pi$ decays by measuring the rate asymmetry between
charge conjugate modes.  The search is performed on a
data sample of $11.1$ million $B\bar{B}$ events recorded on
 the $\Upsilon(4S)$ resonance by the Belle experiment at KEKB.  
We measure  $90\%$
confidence intervals in the partial rate asymmetry ${\cal A}_{\rm CP}$ of  
$-0.25 < {\cal A}_{\rm CP}(K^\mp\pi^\pm) < 0.37$, 
$-0.40 < {\cal A}_{\rm CP}(K^\mp\pi^0) < 0.36$, 
and $-0.53 < {\cal A}_{\rm CP}(K^0\pi^\mp) < 0.82$.  By combining the
$K^\mp\pi^\pm$ and $K^\mp\pi^0$ final states, we conclude that $-0.22 <
{\cal A}_{\rm CP}(K^\mp(\pi^\pm+\pi^0)) < 0.25$ at the $90\%$ confidence
level. 
\end{abstract}

%
\normalsize
%
%
%

The most straightforward indication for CP violation in the $B$ meson
system would be a time-independent rate asymmetry between CP
conjugate decays into flavor specific or self tagging final states. 
Direct CP violation (DCPV)~\cite{bib:hanagaki} of this type will occur in a decay containing
at least two amplitudes that have different CP conserving and CP
violating phases.  The charge asymmetry will be most evident in decays
where the two amplitudes are of comparable strength.  The partial rate
asymmetry can be written as
$${\cal A}_{\rm CP} = {N(\bar{B}\rightarrow \bar{f}) -
N(B\rightarrow f)  \over N(\bar{B}\rightarrow \bar{f}) +
N(B\rightarrow f) } = {2|A_1||A_2|\sin{\delta}\sin{\phi} \over |A_1|^2
+ |A_2|^2 + 2|A_1||A_2|\cos{\delta}\cos{\phi} }, $$
where $\delta$ and $\phi$ are the CP conserving and CP
violating relative phases between amplitudes $A_1$ and $A_2$; $B$
represents either a $B^0_d$ or $B^+$ meson, $f$ represents a self
tagging final state, and $\bar{B}$ and $\bar{f}$ are the conjugate states. 

In the Standard Model, DCPV occurs in charmless hadronic $B$ decay modes that involve both
penguin (P) amplitudes and weak $b\rightarrow u$ tree (T) amplitudes
containing the CP violating weak phase $\phi_3 = \arg{
(V^*_{ub})}$ (in a standard convention~\cite{bib:wolfenstein}). Two
examples are $B^0\rightarrow K^+\pi^-$ and $B^+\rightarrow K^+\pi^0$. 
Uncertainties in the relative size of the P and T amplitudes as well
as their relative strong phase make it difficult to predict the
magnitude of ${\cal A}_{\rm CP}$.  Attempts have been made to
estimate or calculate these
quantities~\cite{bib:neubert0,bib:delphine,bib:sanda1};
however, the different methods give conflicting results. Although
absolute predictions vary, they tend to agree that ${\cal A}_{\rm CP}$ has approximately the same magnitude and
sign in the $K^\mp\pi^\pm$ and $K^\mp\pi^0$ final
states~\cite{bib:neubert0,bib:sanda1,bib:rosner1}.  This suggests that we
can combine these two modes to increase the statistical precision. 

At the quark level, the $B^+\rightarrow K^0\pi^+$ process does not
include a contribution from the tree diagram; thus one expects a negligible
asymmetry in this mode.  DCPV can be induced in this mode through
rescattering effects~\cite{bib:george}; however, these are expected to be small~\cite{bib:neubert0,bib:sanda1,bib:rosner1}.

In this paper, we describe a search for DCPV by
measuring the difference between the yields of $\bar{B}$ and $B$
decays into the self tagging final states $K^\mp\pi^\pm$, $K^\mp\pi^0$,
and $K^0_S\pi^\mp$. Searches for partial rate asymmetries in the
$B\rightarrow K\pi$ decays have been
reported in~\cite{bib:cleo}. DCPV has been
observed in the kaon system~\cite{bib:na48,bib:ktev}.

The analysis is based on data taken by the Belle
detector~\cite{bib:belle} at the KEKB~\cite{bib:kekb}
$e^+e^- $ storage ring.  The data set consists of $10.4$ $\rm fb^{-1}$ on
the $\Upsilon(4S)$ resonance, which corresponds to $11.1 \times 10^6 $ $B\bar{B}$
events. A $0.6$ $\rm fb^{-1}$ data sample is taken $60$ MeV
below the $\Upsilon(4S)$ resonance to perform systematic studies of the
continuum $e^+e^- \rightarrow q\bar{q}$ background.
KEKB is a two-ring, energy-asymmetric storage ring with electrons stored at
$8$ GeV and positrons stored at $3.5$ GeV giving an
$\Upsilon(4S)$ center-of-mass system (cms) boosted by $\gamma\beta =
0.425$ in the electron beam direction.
The Belle detector is described in detail in~\cite{bib:belle}. 
Charged $\pi$ and $K$ mesons are
identified based on their energy loss ($dE/dx$) in the central drift
chamber and their \v{C}erenkov light yield in an array of silica Aerogel
counters (ACC).  For each $K$ and $\pi$ hypothesis, the $dE/dx$ and ACC probability density functions are
combined to form 
 a single likelihood, ${\cal L}_K$ or ${\cal L}_\pi$. $K$ and $\pi$ mesons are distinguished by cutting
on the likelihood ratio ${\cal L}_K/({\cal L}_K + {\cal
L}_\pi)$. The particle identification (PID) efficiency
is determined using nearly pure samples of $K$ and $\pi$ mesons tagged using the
continuum $D^{*+}$ decay chain $D^{*+} \rightarrow D^0\pi^+$, $D^0
\rightarrow K^-\pi^+$~\cite{bib:charge}. For tracks in the
$B\rightarrow hh$ ($h=K$ or $\pi$) signal
region of $2.4$ GeV$/c$ $<p_{\rm  cms}<2.8$ 
GeV$/c$ ($1.5$ GeV$/c$ $<p_{ \rm lab}<4.5$  GeV$/c$) the $K$ efficiency and
fake rate are  $ \epsilon_K = 0.85$ and $f_K = 0.10$ (true $K$
fakes $\pi$);  the $\pi$ efficiency and fake rate are
$\epsilon_\pi = 0.92$ and $ f_\pi = 0.04$ (true $\pi$ fakes $K$).  The
relative systematic error is $2.5\%$ in the efficiencies and $10\%$ in
the fake rates.

In the $K^\mp\pi^\pm$ modes, the asymmetry is diluted by double
mis-identification of $K^+\pi^-$ as $\pi^+K^-$.  The
dilution is given by 
$$ {\cal A}_{\rm CP}^{\rm meas.} = (1-2w){\cal
A}_{\rm CP}^{\rm true},$$
where $w = f/(\epsilon + f)$ is the wrong tag fraction,
$f = f_Kf_\pi = 0.0040 \pm 0.0006$ is the double mis-identification rate and
$\epsilon = \epsilon_K\epsilon_\pi = 0.78 \pm 0.03$ is the PID
efficiency.  This gives a wrong tag fraction of $w = 0.0051 \pm 0.0007$ and a dilution of $1-2w = 0.990 \pm 0.002$. The value of ${\cal
A}_{\rm CP}(K^\mp\pi^\pm)$ is corrected by this factor.

The identification procedure of $B\rightarrow K\pi$ decays has been
documented in~\cite{bib:kazuhito}.  Here we give a brief overview of the
analysis.
 All charged tracks that originate from the interaction point are
considered as kaon or pion candidates.   $K^0_S$ mesons are reconstructed using
pairs of charged tracks that have an invariant mass with
$|m(\pi^+\pi^-)-m(K^0)|<30$ MeV$/c^2$.  The candidate must have a
displaced vertex and flight direction consistent with a high
momentum $K^0_S$ originating from the interaction point. Candidate $\pi^0$
mesons are formed by requiring pairs of $\gamma$s to have
$|m(\gamma\gamma)-m(\pi^0)|< 16$  MeV$/c^2$. 
Continuum background is reduced by cutting on a
likelihood ratio formed from event topology variables and initial state angular
momentum variables such as the $B$ flight direction or the decay axis direction.
This cut removes more than $95\%$ of the continuum background while
retaining $40$ to $50\%$ of the signal depending on the final state.
Further details can be found in~\cite{bib:kazuhito}.

To reconstruct the $B$ meson, we form two quantities: the energy difference, $\Delta
E = E_{B} - E_{\rm beam}$, and the beam constrained mass, $m_{bc} =
\sqrt{E_{\rm beam}^2 - p_B^2}$, where $E_{\rm beam} = \sqrt{s}/2 = 5.29$ GeV, 
and $E_{B}$ and $p_B$ are the reconstructed energy and momentum of the
$B$ candidate in the cms. 
For the $K^\mp \pi^\pm$ and $K^0_S\pi^\mp$ final states, the signal
yield is extracted by fitting the $\Delta E$ distribution in the
region
$m_{bc}>5.27$
 GeV$/c^2$ and $|\Delta E|<250$ MeV.
As the $\Delta E$ distributions for the $K^\mp\pi^0$ final states have
large asymmetric tails, we apply a two-dimensional fit to the
$\Delta E$ and $m_{bc}$ distributions in the region $m_{bc}>5.2$
 GeV$/c^2$ and $-450$ MeV $<\Delta E<150$ MeV.  The Gaussian signal shapes
are calibrated using the $B^+ \rightarrow \bar{D^0}\pi^+$
data sample and high momentum $D^0\rightarrow K^-\pi^+$,
$K^-\pi^+\pi^0$, and $D^+\rightarrow K^0_S\pi^+$ decays where the
daughter particles have the same momenta as signal particles~\cite{bib:charge}.
The continuum background shapes are determined from sideband data.  We use a linear function for the $\Delta E$
background; we parameterize the background for $m_{bc}$ with a
function that behaves like phase space near the endpoint (the ARGUS
shape~\cite{bib:ARGUS}). In all fits, the signal and background yields are the only free parameters.

\begin{figure}[htbp]
\centerline{\epsfysize 4.0 truein\epsfbox{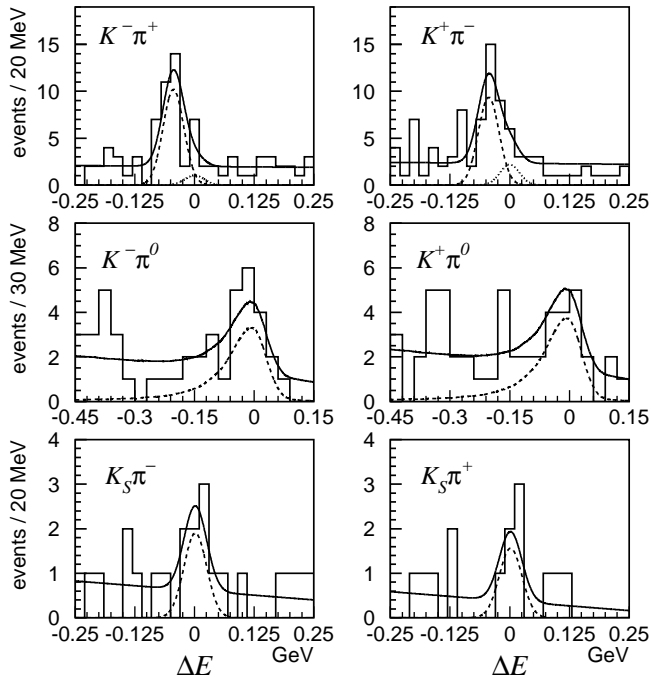} }
\caption{The $\Delta E$  distributions for $\bar{B}$ candidates (left)
and $B$ candidates (right). 
The solid line
represents the sum of signal Gaussian (asymmetric for $K^\pm\pi^0$) and a
linear background function determined from the fit.  The signal component is represented by
the dashed line. In the $K^\mp\pi^\pm$ fits, the signal component is centered at $-44$ MeV due to $\pi$ mass assignment to
the $K$ track.  A component centered at zero (dotted line) is
also included to account for mis-identified
$B^0\rightarrow \pi^+\pi^-$ backgrounds.  The $K^\mp\pi^0$ distributions are projections
of a two-dimensional $m_{bc}$ and $\Delta E$ fit onto the $\Delta E$
distribution in the $m_{bc}$ signal region. }
\label{fig:de}
\end{figure}

Figure~\ref{fig:de} shows the $\Delta E$ distributions for the
$\bar{B}$ and $B$ candidates.  
For $K^\mp\pi^\pm$, the particle energy is calculated assuming a pion
mass which shifts the signal peak to $-44$ MeV.  A component centered
at zero is also
included to account for mis-identified
$B^0\rightarrow \pi^+\pi^-$ backgrounds.  For the $K^\mp \pi^0$ and
$K^0_S\pi^\mp$ distributions, the correct mass assignments are used.
The results are listed in Table~\ref{tab:results}. 
For the
combined $K^\mp\pi$ asymmetry, the number of events is $N(\bar{B}) =
N(K^-\pi^0) + {1\over 1-2w}N(K^-\pi^+)$.
In
all cases, the measured asymmetry is consistent with zero.  The
results are also compared to the 
predictions in~\cite{bib:sanda1} assuming $\phi_3=90^\circ$.  Other
authors~\cite{bib:neubert0} expect asymmetries that are
less than $10\%$.

\begin{table}[htbp]
\caption{Results of searches for DCPV in $B\rightarrow K\pi$
decays. Listed are the number of signal events for each final state,
the ${\cal A}_{\rm CP}$ values with error, their $90\%$ confidence intervals,
and theoretical estimates from Ref. [5].  The first error is statistical; the second is
systematic.  The $90\%$ statistical confidence intervals have been
increased by the systematic error. In the $K^\mp\pi^\pm$ final
states, the asymmetry is corrected for the dilution $(1-2w)$, due to
double mis-identification.  For the combined $K^\mp\pi$ asymmetry, the
number of events are combined as $N(\bar{B}) = N(K^-\pi^0) +
{1\over 1-2w}N(K^-\pi^+)$.}
\begin{center}
\begin{tabular}{l c c c c c}
Mode & $N(\bar{B})$ & $N(B)$ & ${\cal A}_{\rm CP}$ & $90\%$ C.L. & Theory~\cite{bib:sanda1}\\
\hline
$K^\mp\pi^\pm$  & $27.7^{+6.8}_{-6.1}$ & $25.4^{+7.0}_{-6.3}$ & 
$0.044^{+0.186}_{-0.167}\hskip 1mm^{+0.018}_{-0.021}$ & $ -0.25$ : $0.37$ 
 & $-0.19$ \\
$K^\mp\pi^0$  & $16.5^{+5.3}_{-4.7}$ & $18.6^{+5.7}_{-5.0}$ &
$-0.059^{+0.222}_{-0.196}\hskip 1mm^{+0.055}_{-0.017}$ & $-0.40$ : $0.36$
 & $-0.18$ \\
$K^0_S\pi^\mp$  & $5.6^{+3.4}_{-2.7}$ & $4.6^{+2.8}_{-2.2}$ &
$0.098^{+0.430}_{-0.343}\hskip 1mm^{+0.020}_{-0.063}$ & $ -0.53 $ : $0.82$ 
 & $-0.01$\\
$K^\mp(\pi^\pm + \pi^0)$ &$44.5^{+8.7}_{-7.7}$ & $44.2^{+9.1}_{-8.1}$
& $0.003^{+0.142}_{-0.126}\hskip 1mm^{+0.017}_{-0.014}$ & $-0.22$ :
$0.25$ & \\
\end{tabular}
\end{center}
\label{tab:results}
\end{table}

The systematic error in the fitting procedure is determined by
varying the fixed parameters of the signal and background functions within their errors. The
parameters are varied simultaneously for the conjugate final states
and the deviation in ${\cal A}_{\rm CP}$ is taken as the systematic
error.  These deviations are on the order of $\pm 0.01$.  We estimate
the error due to other rare $B$ decay backgrounds
such as $B\rightarrow \rho\pi$ or $K^*\pi$ from the change in ${\cal
A}_{\rm CP}$ when the negative $\Delta E$ region is excluded from the fit.
 These backgrounds are shifted in $\Delta E$ by the mass of the
missing particle, typically $-140$ MeV. This is the
dominant systematic error in each mode: $-0.017$ for $K^\mp \pi^\pm$,
$+0.05$ for $K^\mp\pi^0$, and $-0.06$ for $K^0_S\pi^\mp$. 
All deviations are then added in quadrature.  

\begin{figure}[htbp]
\centerline{\epsfysize 2.0 truein\epsfbox{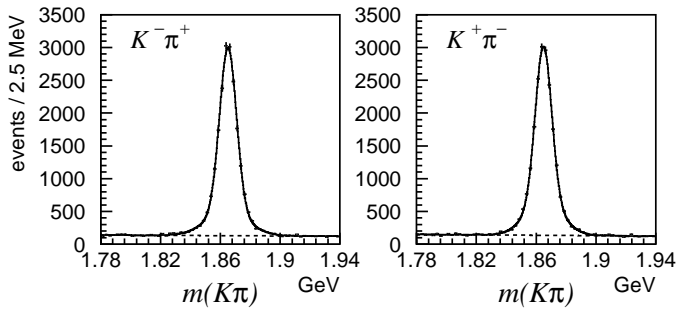} }
\caption{The $K^\pm \pi^\mp$ invariant mass distributions for $D$ meson
candidates.  The $D$ peaks are fitted with a
$\sigma = 5.5$
MeV Gaussian and a second $\sigma = 13.6$ MeV Gaussian containing
$26\%$ of the area.
 The solid curve is the sum the signal Gaussians and a linear
background function (dashed line).  The asymmetry in the yield is $(1
\pm 6)\times 10^{-3}$.}
\label{fig:md1}
\end{figure}

We study the effects of detector-based asymmetries using a sample of
high momentum $D\rightarrow K\pi$ decays.
Figure~\ref{fig:md1} shows mass distributions for $D^0\rightarrow K^-\pi^+$ and
$\bar{D^0}\rightarrow K^+\pi^-$ candidates. The daughter tracks are
required to have $|p_{\rm lab}|>1.5$ GeV$/c$ and the same PID
requirements as signal tracks. The $D$ mass peak is fitted with a double
Gaussian and a linear background function.  We float all parameters in
the fit to test for differences in resolution between the
conjugate states. The signal and background shapes all agree within the errors
for the conjugate states.  We measure $19091\pm 172$
$D^0$ candidates and $19055 \pm 164$ $\bar{D^0}$ candidates giving
an asymmetry of $(1 \pm 6) \times 10^{-3}$. We study
this further using an inclusive sample of $2.2\times 10^6$ charged
tracks ($h^\pm$)~\cite{bib:electrons} required to have the same momentum and angular distributions
as signal tracks.  We measure an asymmetry
of $(N(h^+)-N(h^-))/(N(h^+)+N(h^-)) =  (8.1 \pm 6.7) \times 10^{-4}$, which is
consistent with Monte Carlo expectations.  We
test for asymmetries in the PID performance by applying kaon and
pion selection cuts to the charged hadron sample. The sample contains $1.4 \times 10^6$ identified pions
and $0.7\times 10^6$ identified kaons.  We measure asymmetries of
$(N(\pi^+)-N(\pi^-))/(N(\pi^+)+N(\pi^-)) =  (1.6 \pm 0.9) \times 10^{-3}$ and
$(N(K^+)-N(K^-))/(N(K^+)+N(K^-)) =  (-0.9 \pm 1.1) \times 10^{-3}$.  These
asymmetries result from a small charge dependence in the $dE/dx$
measurement.  Taking these numbers into account, we
conservatively assign a $1\%$ systematic error for detector-based
asymmetries in all modes.  This is added in quadrature
to the fitting systematics discussed above, as well as the error in
the dilution factor where appropriate.  We add the systematic errors
to the statistical $90\%$
confidence intervals to obtain the final $90\%$ confidence intervals
listed in Table~\ref{tab:results}.

In conclusion, we have performed a search for DCPV in
self tagging $B\rightarrow K\pi$ decays. All results are consistent
with a null asymmetry and we conclude at the $90\%$ confidence level
that $-0.25 < {\cal A}_{\rm CP}(K^\mp\pi^\pm) <0.37 $, $-0.40 < {\cal
A}_{\rm CP}(K^\mp\pi^0) <0.36 $, and $-0.53 < {\cal A}_{\rm
CP}(K^0_S\pi^\mp) <0.82 $. For the combined
 $K^\mp\pi^\pm$ and $K^\mp\pi^0$ data samples we determine $ -0.22<
{\cal A}_{\rm CP}(K^\mp(\pi^\pm + \pi^0)) <0.25 $.  These results are consistent with the limits set
in~\cite{bib:cleo}.
The excellent performance of the Belle high momentum PID allows us to
reduce significantly the dilution due to double mis-identification.
This will become a crucial factor as the Belle data samples increase significantly.

We wish to thank the KEKB accelerator group for the excellent
operation of the KEKB accelerator. We acknowledge support from the
Ministry of Education, Culture, Sports, Science, and Technology of Japan
and the Japan Society for the Promotion of Science; the Australian Research
Council and the Australian Department of Industry, Science and
Resources; the Department of Science and Technology of India; the BK21
program of the Ministry of Education of Korea and the CHEP SRC
program of the Korea Science and Engineering Foundation; the Polish
State Committee for Scientific Research under contract No.2P03B 17017;
the Ministry of Science and Technology of Russian Federation; the
National Science Council and the Ministry of Education of Taiwan; the
Japan-Taiwan Cooperative Program of the Interchange Association; and
the U.S. Department of Energy.

\onecolumn


\begin{thebibliography}{99}
\bibitem{bib:hanagaki}{In neutral $B$ decays, this asymmetry can also
be induced by CP violation in mixing; however, this effect is small.
See for instance OPAL Collaboration, K. Ackerstaff {\it et al.},
Z. Phys. C {\bf 76}, 401 (1997).}
\bibitem{bib:wolfenstein}{L. Wolfenstein, Phys. Rev. Lett. {\bf 51}, 1945 (1983).}
\bibitem{bib:neubert0}{M. Beneke, G. Buchalla, M. Neubert,
C.T. Sachrajda, Nucl. Phys. B {\bf 591} 313 (2000), and hep-ph/0104110.} 
\bibitem{bib:delphine}{ D. Delepine, J.M. Gerard, J. Pestieau,
J. Weyers, Phys. Lett. B {\bf 429}, 106 (1998).}
\bibitem{bib:sanda1}{Y.-Y. Keum, H.-N. Li, A.I. Sanda, Phys. Rev. D
{\bf 63}
054008 (2001).} 
\bibitem{bib:rosner1}{M. Gronau, J. L. Rosner, Phys. Rev. D {\bf 59}
113002 (1999).}
\bibitem{bib:george}{W.S. Hou, K.C. Yang, Phys. Rev. Lett. {\bf 84}, 4806 (2000).}
\bibitem{bib:cleo}{CLEO Collaboration, S. Chen {\it et al.},
Phys. Rev. Lett. {\bf 85}, 525 (2000).}
\bibitem{bib:na48}{NA48 Collaboration, V. Fanti {\it et al.},
Phys. Lett. B {\bf 465}, 335 (1999).}
\bibitem{bib:ktev}{KTeV Collaboration, A. Alavi-Harati {\it et al.},
Phys. Rev. Lett. {\bf 83}, 22 (1999).}
\bibitem{bib:belle}{ Belle Collaboration, K. Abe {\it et al.}, KEK Progress Report 2000-4, 
submitted to Nuc. Inst. Meth. A.}
\bibitem{bib:kekb} {KEK B-Factory Design
Report KEK-Report-95-7 (1995) (unpublished).}
\bibitem{bib:charge} {For these calibration samples, inclusion of
charge conjugate modes is implied.}
\bibitem{bib:kazuhito}{Belle Collaboration, K. Abe {\it et al.}, Belle Preprint 2001-5, hep-ex/0104030, submitted to Phys. Rev. Lett.}
\bibitem{bib:ARGUS} {ARGUS Collaboration, H. Albrecht {\it et al.},
Phys. Lett. B {\bf 241}, 278
(1990).} 
\bibitem{bib:electrons} {Electrons and positrons are removed using
calorimeter and $dE/dx$ information.}
\end{thebibliography}
\end{document}